\documentclass{Interspeech}

\interspeechcameraready

\usepackage{comment}
\usepackage{tabularx}
\usepackage{pifont}
\usepackage{subfig}
\usepackage{adjustbox}
\usepackage{multirow}
\usepackage{multicol}

\usepackage[dvipsnames, table, xcdraw]{xcolor} 
\usepackage{pifont} 
\newcommand{\cmark}{\ding{51}}%
\definecolor{bsRed}{rgb}{0.95, 0.0, 0.0}
\usepackage{hyperref}
\definecolor{bsRed}{rgb}{0.95, 0.0, 0.0}
\definecolor{mygray}{RGB}{128,128,128} 
\definecolor{lightgray}{RGB}{200,200,200}
\usepackage{hyperref}
\usepackage{cite}

\begin{document}

\title{Codec-Based Deepfake Source Tracing via Neural Audio Codec Taxonomy
\thanks{*Equal Contribution. $\dagger$ Corresponding Author}
}

\author[affiliation={1*}]{Xuanjun}{Chen}
\author[affiliation={2*}]{I-Ming}{Lin}
\author[affiliation={4}]{Lin}{Zhang}
\author[affiliation={2}]{Jiawei}{Du}
\author[affiliation={5^{\dagger}}]{Haibin}{Wu}
\authorbreak
\author[affiliation={1,3}]{Hung-yi}{Lee}
\author[affiliation={2}]{Jyh-Shing Roger}{Jang}

\affiliation{}{Graduate Institute of Communication Engineering}{National Taiwan University}
\affiliation{}{Department of Computer Science and Information Engineering}{National Taiwan University}
\affiliation{}{Department of Electrical Engineering}{National Taiwan University}
\affiliation{}{Speech@FIT}{Brno University of Technology \quad $^5$Independent Researcher}

\keywords{Anti-spoofing, source tracing, audio deepfake detection, neural audio codec, explainability}

\maketitle

\begin{abstract}
Recent advances in neural audio codec-based speech generation (CoSG) models have produced remarkably realistic audio deepfakes. We refer to deepfake speech generated by CoSG systems as codec-based deepfake, or CodecFake.
Although existing anti-spoofing research on CodecFake predominantly focuses on verifying the authenticity of audio samples, almost no attention was given to tracing the CoSG used in generating these deepfakes. 
In CodecFake generation, processes such as speech-to-unit encoding, discrete unit modeling, and unit-to-speech decoding are fundamentally based on neural audio codecs.
Motivated by this, we introduce 
source tracing for CodecFake via neural audio codec taxonomy, which dissects neural audio codecs to trace CoSG.
Our experimental results on the \textit{CodecFake+} dataset provide promising initial evidence for the feasibility of CodecFake source tracing while also highlighting several challenges that warrant further investigation. 
\end{abstract}
\section{Introduction}  
Deepfake detection distinguishes bona fide audio (spoken by humans) from deepfakes (generated by machines). 
Numerous challenges and datasets~\cite{li2024audio,wu2023defender} have been created to drive research in this field, e.g., ASVspoof \cite{wu15e_interspeech, kinnunen17_interspeech, todisco2019asvspoof, Liu_2023, Wang2024_ASVspoof5} and ADD \cite{yi2022add, yi2024add2023}. 
Beyond detecting deepfake speech, tracing its source has recently gained attention. 
Source tracing can provide valuable cues to identify the origin or creator of deepfake audio, which is particularly useful in out-of-domain deepfake detection. 
The track 3 of ADD 2023 \cite{yi2024add2023} focused on this topic to recognize the algorithms of the deepfake utterances. 
In Interspeech 2025, a special session\footnote{\href{https://deepfake-total.com/sourcetracing}{https://deepfake-total.com/sourcetracing}} is being held to encourage researchers to develop source tracing techniques for deepfake speech. 

The most straightforward way of source tracing \cite{borrelli2021synthetic} is tracing deepfake speech back to its originating attack model ID \cite{yi2024add2023}. 
To improve generalization, researchers start working more flexibly, known as attribute-based source tracing \cite{Yan2022AnII, Zhang2023DistinguishingNS, zhu2022source, klein24_interspeech}. 
Instead of using deepfake system IDs as categories, it categorizes deepfake systems based on their attributes, given that some deepfake systems share common components. 
This enables better generalization to trace spoofing algorithms not seen during training but composed of known building blocks \cite{klein24_interspeech}.
Guided by well-defined attributes in conventional text-to-speech (TTS) and voice conversion (VC) systems, researchers in source tracing pay their attention to classifying attributes: input types \cite{klein24_interspeech}, acoustic models \cite{Zhang2023DistinguishingNS, klein24_interspeech}, speaker representation \cite{zhu2022source}, and vocoders \cite{zhu2022source, Yan2022AnII, Zhang2023DistinguishingNS, klein24_interspeech}, etc.

However, all the aforementioned studies focus on source tracing for conventional TTS/VC, whereas tracing sources in the emerging codec-based speech generation (CoSG) models remains largely unexplored. 
For the new CodecFake \cite{wu24p_interspeech, xie2025codecfake, chen2025codecfake+}, source tracing was preliminarily explored in \cite{xie2025neural}. 
They treat the source tracing task as a closed-set classification problem for category deepfake systems ID. 
Although they introduced an additional binary out-of-distribution classification, it cannot provide deeper insights when encountering multiple unseen CoSG systems. 
Overall, source tracing for CodecFake remains limited and urgently requires exploration.

Recently, \textit{CodecFake+} \cite{chen2025codecfake+}, a large-scale CodecFake database, was proposed.
It covers most currently available neural audio codecs and CoSG systems.
The dataset and its well-structured taxonomy serve as valuable reference guides for source tracing.
Based on \textit{CodecFake+}, we defined taxonomy-based source tracing tasks for CodecFake speech, including vector quantization classification, auxiliary objective classification, and decoder type classification.
Additionally, we explored different training strategies and taxonomy-guided balanced training data selection. 
Our experiments show that a multi-task framework incorporating the three classification tasks enhances source tracing performance. Additionally, selecting balanced training data guided by auxiliary objectives shows potential for improving the generalization of source tracing. We will release our entire codebase to facilitate further research\footnote{\href{https://github.com/ResponsibleGenAI/CodecFake-Source-Tracing}{https://github.com/ResponsibleGenAI/CodecFake-Source-Tracing}}. 

\section{Neural Audio Codec Taxonomy}

Neural audio codecs \cite{arora2025landscape, wu2024towards, wu2024codec, wu2024codec_slt24} have emerged as not only a core technology to compress and transmit audio signals efficiently, but also a core component for high-quality speech generation systems.
Yet, unlike traditional TTS/VC pipelines with clearly delineated acoustic models and vocoders, codec-based speech generation systems often lack a standardized, transparent structure. 
This obstacle complicates detecting and tracing codec-based deepfake audio to its true source. 
A recent work proposed a large-scale \textit{CodecFake+} with a comprehensively defined taxonomy~\cite{chen2025codecfake+}, such as vector quantization, auxiliary objectives, and decoder types. 
From this, we gain a structured lens to understand and categorize various codec methods. 
The taxonomy is categorized into three groupings as follows:

\subsection{Vector Quantization (VQ)}
Vector quantization methods aim to map frame-wise audio features to discrete tokens. The VQ grouping categorization can have three sub-categories:
\begin{itemize}
    \item \textbf{Multi-codebook Vector Quantization (Mvq)}: Multiple quantizers are applied sequentially to refine the residuals of previous ones. 

    \item \textbf{Single-codebook Vector Quantization (Svq)}: Uses a single codebook to encode fewer tokens, reducing token usage. 

    \item \textbf{Scalar-codebook Vector Quantization (Scq)}: 
    Mapping complex signals into a compact latent space based on a scalar codebook. 
\end{itemize}

\subsection{Auxiliary Objectives (AUX)}
Auxiliary objectives provide additional supervision to guide the training of neural audio codecs. 
Three sub-categories are:
\begin{itemize}
    \item \textbf{Semantic Distillation (Sem)}: Enhances codec representation with semantic information.

    \item \textbf{Disentanglement (Disent)}: Separates speech attributes (e.g., content, prosody, timbre) into distinct representations. 
    
    \item \textbf{None}: refer to no auxiliary Objectives.
\end{itemize}

\subsection{Decoder Types (DEC)}
The decoder aims to reconstruct waveforms from discrete units, with upsampling being a critical operation. There are two main methods based on the processing domain:
\begin{itemize}
    \item \textbf{Time Decoder (Time)}: Uses transposed convolution to upsample codec representations into raw audio waveforms. 

    \item \textbf{Frequency Decoder (Freq)}: Uses Inverse Short-Time Fourier Transform to upsample decoded features. 

\end{itemize}

In summary, the taxonomy defined above underlie nearly all neural audio codec designs. 
Thus, they serve as robust indicators of the ``fingerprint'' of a neural audio codec, enabling the tracing of CoSG systems based on their codec components, even when applied to out-of-domain CoSG systems.

\section{Codec-Based Deepfake Source Tracing}

\subsection{Task Definition}\label{sec:task_definition}

\begin{figure}[t]
    \centering
    \includegraphics[width=1\columnwidth]{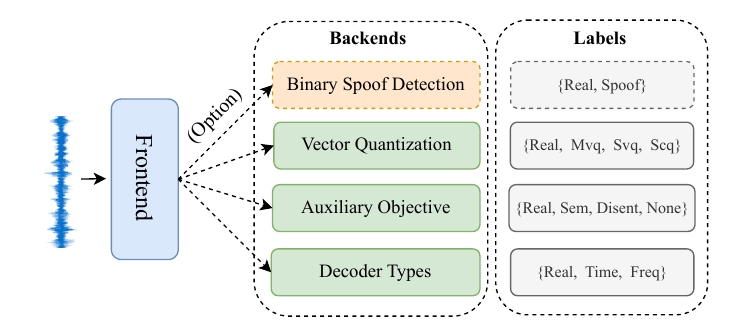}
    \vspace{-2em}
    \caption{A multi-task source tracing framework.}
    \label{fig:multi-task model}
\end{figure}

Building on the neural audio codec taxonomy, we define three multi-class classification tasks for CodecFake source tracing:
\begin{itemize}
    \item \textbf{Vector Quantization Classification (VQ Task)}: Identifies the codec's vector quantization approach of a given speech.
    \item \textbf{Auxiliary Objective Classification (AUX Task)}: Classifies the codec's auxiliary objective of a given speech.
    \item \textbf{Decoder Type Classification (DEC Task)}: Classifies the codec's decoder type of a given speech.
\end{itemize}
Additionally, we include a traditional anti-spoof detection task:
\begin{itemize}
    \item \textbf{Binary Spoof Detection (BIN Task)}: Determines whether a given audio is bona fide or spoofed speech. 
\end{itemize}
A detailed description of the labels for each multi-class classification task is provided in Figure~\ref{fig:multi-task model}.

\subsection{Multi-Task Training for Source Tracing}\label{model_architecture}
As shown in Figure~\ref{fig:multi-task model}, the waveform is first fed into a front-end to convert to a hidden presentation. 
It is then processed by four independent backends, each having a similar structure but producing outputs in different dimensions for its corresponding task, as introduced in the previous subsection. 
During training, the output score from each backend is used to compute its corresponding loss. 
Then, the total loss is obtained by a weighted sum as follows:
\begin{equation}
\label{tab:multi-task_loss}
l_{\text{total}} 
= \lambda_1\,l_{\text{BIN}} 
+ \lambda_2\,l_{\text{VQ}} 
+ \lambda_3\,l_{\text{AUX}} 
+ \lambda_4\,l_{\text{DEC}}.
\end{equation}
Here, 
$l_{\text{BIN}}, l_{\text{VQ}}, l_{\text{AUX}}, l_{\text{DEC}}$ represent the losses for binary spoof detection, vector quantization classification, auxiliary objective classification, and decoder type classification, respectively. 
Let \([\lambda_1, \lambda_2, \lambda_3, \lambda_4]\) denote the weights for each subtask. When a subtask is added during training, its corresponding \(\lambda_i\) is included; otherwise, it is omitted. 
The study explores different training configurations, detailed in Section \ref{sec:exp_setup}, and summarized in Table~\ref{tab:table_train_config}. 

\begin{table*}[t]
\caption{Summary of Source Tracing Dataset Based on \textit{CodecFake+} \cite{chen2025codecfake+}. The BIN represents the overall CodecFake dataset statistics used in this paper, while Spoof (VQ, AUX, DEC) provides detailed statistics of the spoofed data from a taxonomy perspective.}
\label{tab:codec-fake_omini_info}
\centering
\fontsize{7}{9}\selectfont
\setlength\tabcolsep{4.5pt}
\vspace{-1em}
\begin{tabular}{lcccccccccccc}
\toprule
& & \multicolumn{2}{c}{BIN} & \multicolumn{3}{c}{Spoof (VQ)} & \multicolumn{3}{c}{Spoof (AUX)} & \multicolumn{2}{c}{Spoof (DEC)}  & \multirow{2}{*}{\#. of Codec}  \\
\cmidrule(r){3-4} \cmidrule(r){5-7} \cmidrule(r){8-10} \cmidrule(r){11-12}
Dataset & Sampling Strategies   & Bona fide         & Spoof  & Mvq & Svq & Scq  & None & Sem & Disent  & Time & Freq     &       \\
\midrule
 CoRS Train      & \multirow{2}{*}{VQ balanced sampling}  &    42,965 & 42,965 & 14,322& 14,322& 14,321 &  32,529  & 4,590& 5,846& 23,245& 19,720&  31  \\
CoRS Validation  &     &    735    &   735  & 245   & 245   & 245   &  559     &  72   & 104   & 403   & 332   &   31  \\

\midrule
CoRS Train  &  \multirow{2}{*}{AUX balanced sampling}  &    42,965 & 42,965 & 36,152& 5,278& 1,535 & 14,322  & 14,321& 14,322& 28,445& 14,520&   31  \\
CoRS Validation  &     &    735    &   735  &  624   & 80   & 31   &  245     & 245   & 245   & 469   & 266   &  31  \\

\midrule
 CoRS Train   &  \multirow{2}{*}{DEC balanced sampling}   &    42,965 & 42,965 & 34,156& 5,926& 2,883 & 25,698  & 10,941& 6,326 & 21,483& 21,482&   31  \\
 CoRS Validation  &  &    735    &   735  & 580   & 90   & 65   &  435     & 179   & 121  & 368   & 367   &  31  \\

\midrule
   CoRS Test  &  \multirow{3}{*}{\textcolor{lightgray}{N/A}}  &    755    & 22,650 & 18,120& 3,020 & 1,510 &  12,835  & 6,040 & 3,775 & 13,590& 9,060 &  31 \\
  CoSG Test (kn. codec) & &    259    &   318  & 249   & 6    & 63    & 275     & 5     & 38    & 240   & 78   &   6   \\
  CoSG Test (All) & &    866    &   930  & 848   & 20    & 63    &  873     & 5     & 52    & 340   & 590   &  9   \\
\bottomrule
\end{tabular}
\end{table*}

\begin{table}[t]
\caption{Training Configurations. The S* notation represents single-task training. For example, S1 refers to two separate models trained independently on BIN and VQ tasks. }
\label{tab:table_train_config}
\centering
\fontsize{7}{9}\selectfont
\setlength{\tabcolsep}{3.5pt}
\vspace{-1em}
\scalebox{0.98}{
\begin{tabular}{c|c|cccc|c}
\toprule
\textbf{Model} & \textbf{Dataset}& \textbf{BIN} & \textbf{VQ} & \textbf{AUX} & \textbf{DEC} & \textbf{Weight} \(W\) \\
\midrule


\multirow{2}{*}{S1} & \phantom{X}VQ balance  &  {\color{Green} \cmark} &  &  &  &  (1.0)  \\
            & \phantom{X}VQ balance  &  &  {\color{Green} \cmark}   &  &  &  (1.0)  \\

\midrule  
\multirow{2}{*}{S2} & AUX balance & {\color{Green} \cmark} &  &  &  & (1.0) \\
            & AUX balance & &  &  {\color{Green} \cmark}  &  &  (1.0)  \\
\midrule
\multirow{2}{*}{S3} & DEC balance &  {\color{Green} \cmark} &  &  &  &  (1.0)   \\
            & DEC balance & &  &  &  {\color{Green} \cmark}  &  (1.0)  \\
\midrule
D1 & \phantom{X}VQ balance  &{\color{Green} \cmark} &  {\color{Green} \cmark}  &  &  & (0.5, 0.5) \\
D2 & AUX balance &{\color{Green} \cmark} &   & {\color{Green} \cmark} &  & (0.5, 0.5) \\
D3 & DEC balance &{\color{Green} \cmark} &  &  & {\color{Green} \cmark}  & (0.5, 0.5) \\
\midrule
M1 & VQ / AUX / DEC & {\color{Green} \cmark} & {\color{Green} \cmark} & {\color{Green} \cmark} & {\color{Green} \cmark} & (0.25, 0.25, 0.25, 0.25) \\ 
\midrule
M2 & VQ / AUX / DEC &   & {\color{Green} \cmark} & {\color{Green} \cmark} & {\color{Green} \cmark} &  (0.3, 0.4, 0.3)   \\
\bottomrule
\end{tabular}
}
\end{table}

\subsection{Inference for Source Tracing and Anti-Spoofing}
For source tracing, each multi-class backend predicts the most probable class based on the output probabilities.
For spoof detection, if the binary backend is included, its score is used directly; otherwise, the bona fide probabilities from multiple backends are multiplied, and a root transformation inspired by prior work~\cite{zhu2022source} is applied. These two inference configurations are M1 and M2, respectively, as summarized in Table~\ref{tab:table_train_config}.

Since M2 was not directly optimized for binary spoof detection, we use the cubic root of the probability product as a heuristic, assuming that the VQ, AUX, and DEC tasks provide complementary cues. 
Based on this heuristic, when training with only the three multi-class source tracing tasks (i.e., VQ, AUX, and DEC tasks), the inference process computes the cubic root of the product of the three bona fide probabilities: 
\vspace{-0.09 in}
\begin{equation} 
    P^\text{bonafide}=\sqrt[3]{P^\text{bonafide}_\text{VQ} \times P^\text{bonafide}_\text{AUX} \times P^\text{bonafide}_\text{DEC}},
\end{equation}
where $P^\text{bonafide}_\text{task}$ denotes the bona fide probability predicted by the corresponding backend.

\section{Experimental Setup}\label{sec:exp_setup}

We investigate source tracing in \textit{CodecFake+} \cite{chen2025codecfake+} dataset, which comprises two subsets: CoRS (speech resynthesized by pre-trained neural audio codec models) and CoSG (speech generated by codec-based speech generation models).  
The original CoRS includes 31 codecs used to resynthesize spoofed audio from VCTK~\cite{yamagishi2019cstr}. The full training dataset consists of 31 × 42,965 spoofed samples. 
However, due to limited computing resources, we performed taxonomy-guided balanced sampling following \textit{CodecFake+}, selecting part of the spoofed samples to maximize data diversity while maintaining efficiency. 
Rather than focusing on CodecFake detection as in \textit{CodecFake+},  our study targets source tracing.
The final training dataset consists of 42,965 bona fide and 42,965 spoofed audio samples.
The CoRS and CoSG test data are collected from 17 CoSG models from the \textit{CodecFake+} dataset, primarily sourced from official demo pages. This setup better reflects real-world scenarios where spoofed data is generated using unknown systems.
Table~\ref{tab:codec-fake_omini_info} summarizes the dataset statistics for source tracing.

We use Wav2Vec2-AASIST~\cite{tak2022automatic} \footnote{\href{https://github.com/TakHemlata/SSL_Anti-spoofing}{https://github.com/TakHemlata/SSL\_Anti-spoofing}}
 as the model backbone, with modifications to fit multi-task learning as shown in Figure~\ref{fig:multi-task model}. 
The pre-trained Wav2Vec 2.0\footnote{\href{https://huggingface.co/facebook/wav2vec2-xls-r-300m}\url{https://huggingface.co/facebook/wav2vec2-xls-r-300m}} \cite{baevski2020wav2vec} and RawNet2 \cite{9414234} Encoder are shared across multiple tasks. 
All inputs are raw waveforms of approximately 4 seconds at a sample rate of 16 kHz, with RawBoost \cite{tak2022rawboost} applied for data augmentation. 
We train all models on an NVIDIA RTX 4070 GPU with a batch size of 8. The initial learning rate is $1\times10^{-6}$, and weight decay is $1\times10^{-4}$. Cross-entropy losses are used for all backends. 

Table~\ref{tab:table_train_config} summarizes the training configurations, where S represents \textbf{S}ingle-task Learning, D represents \textbf{D}ual-task Learning with BIN and one out of VQ/AUX/DEC, and M represents \textbf{M}ulti-task learning.
We use the Equal Error Rate (EER\%) for deepfake detection as the evaluation metric, where lower values indicate better performance. We use the sample-weighted F1 score for three source-tracing sub-tasks to assess the performance, where higher values indicate better performance.

\section{Experiment Results}
\begin{table}[t]
\centering
\fontsize{7}{9}\selectfont
\setlength\tabcolsep{4pt}
\caption{Results of evaluation on different source tracing tasks.}
\label{tab:exp_result_vq_aux_dec}
\vspace{-1em}
\begin{minipage}{0.41\textwidth}
\centering
\textbf{(a) Vector Quantization Classification}
\vspace{0.5em}
\begin{tabularx}{\textwidth}{@{}ccccccc@{}}
\toprule
\multicolumn{1}{c}{Model} & \multicolumn{2}{c}{CoRS} & \multicolumn{2}{c}{CoSG (kn. codec)} & \multicolumn{2}{c}{CoSG (All)} \\
       \cmidrule(lr){2-3}   \cmidrule(lr){4-5}  \cmidrule(lr){6-7}
 (in Table \ref{tab:table_train_config})  & EER $\downarrow$  & F1 $\uparrow$ & EER $\downarrow$ & F1  $\uparrow$      & EER $\downarrow$ & F1  $\uparrow$ \\
\midrule
Random  & 50.28       & 32.26      & 50.82          & 29.21          & 49.97          & 30.79         \\
\midrule
S1                   & 2.78        & \textbf{96.89} & \textbf{21.66}          & 42.94          & 21.93          & 35.26         \\
D1                   & 2.65        & 95.92          & 24.27          & 45.55          & \textbf{15.58}          & 38.78         \\
M1                   & \textbf{2.62} & 96.15        & 26.69          & 46.51          & 17.20         & 46.35         \\
M2                   & 2.78        & 96.27          & 40.90          & \textbf{48.14} & 49.08          & \textbf{49.60 }       \\
\bottomrule
\end{tabularx}
\end{minipage}%

\begin{minipage}{0.41\textwidth}
\centering
\textbf{(b) Auxiliary Objective Classification}
\vspace{0.5em}
\begin{tabularx}{\textwidth}{@{}ccccccc@{}}
    \toprule
    \multicolumn{1}{c}{Model} & \multicolumn{2}{c}{CoRS} & \multicolumn{2}{c}{CoSG (kn. codec)} & \multicolumn{2}{c}{CoSG (All)} \\
           \cmidrule(lr){2-3}   \cmidrule(lr){4-5}  \cmidrule(lr){6-7}
          (in Table \ref{tab:table_train_config})  & EER $\downarrow$  & F1 $\uparrow$ & EER $\downarrow$ & F1  $\uparrow$      & EER $\downarrow$ & F1  $\uparrow$ \\
    \midrule
    Random               & 50.28       & 28.78      & 50.82          & 30.94          & 49.97          & 31.04         \\
    \midrule
    S2                   & 2.77        & 96.63     & \textbf{22.87} & 32.80   & 14.91          & 19.44         \\
    D2                   & 1.62    & \textbf{97.25} & 27.74          & 34.08          & 16.97          & 19.12         \\
    M1                   & \textbf{1.32}       & 96.71      & 33.62          & 31.71          & \textbf{14.41}          & 15.85         \\
    M2                & 1.59  & 96.65      & 33.62       & \textbf{40.00}    & 29.33         & \textbf{26.78} \\
    \bottomrule
\end{tabularx}
\end{minipage}%

\begin{minipage}{0.41\textwidth}
\centering
\textbf{(c) Decoder Types Classification}
\vspace{0.5em}
\begin{tabularx}{\textwidth}{@{}ccccccc@{}}
\toprule
\multicolumn{1}{c}{Model} & \multicolumn{2}{c}{CoRS} & \multicolumn{2}{c}{CoSG (kn. codec)} &     \multicolumn{2}{c}{CoSG (All)} \\
    \cmidrule(lr){2-3}   \cmidrule(lr){4-5}  \cmidrule(lr){6-7}
 (in Table \ref{tab:table_train_config}) & EER $\downarrow$  & F1 $\uparrow$ & EER $\downarrow$ & F1  $\uparrow$      & EER $\downarrow$ & F1  $\uparrow$ \\
\midrule
Random & 50.28       & 38.97      & 50.82       & 36.89      & 49.97         & 34.37         \\
\midrule
S3                   & 2.38        & \textbf{97.50}   & \textbf{24.78}  & 41.54  & \textbf{11.91} & \textbf{27.16}  \\
D3                   & 1.57        & 95.00      & 29.29    & 39.44  & 34.67          & 16.56         \\
M1                   & \textbf{1.16}       & 95.94      & 33.97          & 40.36          & 19.42          & 15.74        \\
M2                & 1.21  & 95.91 & 32.41          & \textbf{45.97}         & 32.78          & 23.15      \\  
\bottomrule
\end{tabularx}
\end{minipage}
\vspace{-3mm}
\end{table}

\subsection{Results of Source Tracing}

We present performance across three source tracing tasks (defined in Section~\ref{sec:task_definition}), with results in Table~\ref{tab:exp_result_vq_aux_dec}. Performance on spoof detection and source tracing is included to investigate their complementary relationship.
Each sub-table corresponds to models trained using the respective balanced sampling strategies and evaluated on the corresponding tasks.
Each row represents a different model. For S* models, performance is reported from two separate models: EER is measured using the one trained on BIN task , while F1 is measured using the one trained on VQ/AUX/DEC. From D* to M2, each row corresponds to a single model evaluated across different datasets.
We also include the random guess as the baseline.
Each column represents a different evaluation subset. 

CoRS evaluation set shows the performance when 
training and evaluation data are re-synthesized using the same neural audio codecs. 
For deepfake detection, EER decreases when introducing more backend branches, from single-task and dual-task setups to multi-task learning, boosting overall performance. 
For source tracing, across the three multi-class classification tasks, all models achieve F1 scores of roughly 96\%–97\%.

CoSG (kn. codec) set presents performance when evaluating only the CoSG data that involves the known codec. In this set, the best EERs for VQ, AUX, and DEC tasks reach 21.66\%, 22.87\%, and 24.78\%, respectively. 
These are about an order of magnitude worse than the CoRS evaluation. Multi-task learning does not appear to help deepfake detection here, though it improves VQ, AUX, and DEC source tracing. 

The CoSG (All) set comprises all CoSG evaluation data.
Compared to CoSG (kn. codec), the best deepfake detection EER improves to 15.58\%, 14.41\%, and 11.91\% across the three sampled datasets.
For source tracing, the multi-task M2 model outperforms or approaches dual-task and single-task approaches, though the decoder-types task achieves only a 23.15\% F1 score, which is slightly lower than the 27.16\% of S3 model.
Source-tracing tasks generally benefit deepfake detection only within the CoRS evaluation set. 
Meanwhile, the M2 model delivers competitive performance with the best model in VQ task and AUX task for both CoSG (kn. codec) and CoSG (All).

Although multi-task learning benefits VQ and AUX source tracing tasks, model performance declines as the mismatch increases from CoRS to CoSG (kn. codec) to CoSG (All). This trend can be explained by the codec-based deepfake generation process, which involves two steps: (1) a neural codec encodes speech into discrete units, and (2) a generative model processes these units before a codec decoder generate speech. Based on the above context, we have two reasons to speculate:
Firstly, when trained on CoRS data from a specific codec, the model performs better on CoSG (kn. codec), where only the generative modeling step differs. This suggests that generative modeling significantly impacts source tracing performance.
Secondly, performance further declines in CoSG (All) due to the use of unknown codecs (e.g., Mel-VAE~\cite{kim2024clam_mel-vae}, SoundStream~\cite{zeghidour2021soundstream}, Single-Codec~\cite{li2024single}, DAC-Vocos~\cite{wang2024maskgct}) in the corresponding CoSG system \cite{kim2024clam_mel-vae, shen2024naturalspeech, li2024single, wang2024maskgct}, adding another variability.
Our findings highlight that generative modeling and unseen codecs influence CodecFake source tracing.

\subsection{The Impact of Various Data Balance Strategies}

We train the model on three taxonomy-guided balanced training sets (listed in Table \ref{tab:codec-fake_omini_info}) to examine how various data balancing strategies affect source tracing. The results are in Table~\ref{tab:data_balance}. Although the DEC-balanced dataset-trained D3 performs slightly better on the DEC task, AUX-balanced dataset-trained models perform well among the three tasks. 
When trained on the AUX‑balanced dataset, M2 achieved the highest F1 on the AUX task (40.00\%) and delivered competitive F1 scores of 48.14\% on VQ and 45.97\% on DEC.
These results indicate that model performance is susceptible to variations in training data distributions. Our experiments suggest that balancing the training sets according to auxiliary objectives yields the strongest generalization on source tracing tasks.

\begin{table}[t]
\centering
\caption{Comparison of different data balance strategies. F1 scores are evaluated on CoSG (kn. codec) subset. (Darker cell color indicates better performance)}
\label{tab:data_balance}
\vspace{-1em}
\fontsize{7}{9}\selectfont
\setlength{\tabcolsep}{4.5pt}
\begin{tabular}{ccccc}
\toprule
        &    & VQ Task (F1 $\uparrow$)  & AUX Task (F1 $\uparrow$)  & DEC Task (F1 $\uparrow$)  \\
\midrule
     \multicolumn{2}{r}{Random}  &  29.21  & 30.94  & 36.89 \\

\midrule
\multirow{4}{*}{\begin{tabular}[c]{@{}c@{}}VQ \\
balanced \\ sampling \end{tabular}}   
    & S1 & \cellcolor[rgb]{0.73,0.73,0.73} 42.94 & \textcolor{lightgray}{N/A} & \textcolor{lightgray}{N/A}\\
    & D1 & \cellcolor[rgb]{0.68,0.68,0.68} 45.55 & \textcolor{lightgray}{N/A} & \textcolor{lightgray}{N/A}\\
    & M1 & \cellcolor[rgb]{0.60,0.60,0.60} 49.75 & \cellcolor[rgb]{0.95,0.95,0.95} 31.78 & \cellcolor[rgb]{0.82,0.82,0.82} 38.23\\
    & M2 & \cellcolor[rgb]{0.66,0.66,0.66} 46.84 & \cellcolor[rgb]{1.00,1.00,1.00} 28.95 & \cellcolor[rgb]{0.88,0.88,0.88} 35.25\\
\midrule
\multirow{4}{*}{\begin{tabular}[c]{@{}c@{}}AUX\\ 
balanced \\ sampling \end{tabular}}
    & S2 & \textcolor{lightgray}{N/A} & \cellcolor[rgb]{0.93,0.93,0.93} 32.80 & \textcolor{lightgray}{N/A}\\
    & D2 & \textcolor{lightgray}{N/A} & \cellcolor[rgb]{0.90,0.90,0.90} 34.22 & \textcolor{lightgray}{N/A}\\
    & M1 & \cellcolor[rgb]{0.66,0.66,0.66} 46.51 & \cellcolor[rgb]{0.95,0.95,0.95} 31.71 & \cellcolor[rgb]{0.78,0.78,0.78} 40.36\\
    & M2 & \cellcolor[rgb]{0.63,0.63,0.63} 48.14 & \cellcolor[rgb]{0.79,0.79,0.79} 40.00 & \cellcolor[rgb]{0.67,0.67,0.67} 45.97\\
\midrule
\multirow{4}{*}{\begin{tabular}[c]{@{}c@{}}DEC\\
balanced \\ sampling \end{tabular}}
    & S3 & \textcolor{lightgray}{N/A} & \textcolor{lightgray}{N/A} & \cellcolor[rgb]{0.67,0.67,0.67} 46.73\\
    & D3 & \textcolor{lightgray}{N/A} & \textcolor{lightgray}{N/A} & \cellcolor[rgb]{0.63,0.63,0.63} 47.94\\
    & M1 & \cellcolor[rgb]{0.62,0.62,0.62} 48.54 & \cellcolor[rgb]{0.93,0.93,0.93} 32.49 & \cellcolor[rgb]{0.77,0.77,0.77} 40.85\\
    & M2 & \cellcolor[rgb]{0.65,0.65,0.65} 47.17 & \cellcolor[rgb]{0.93,0.93,0.93} 32.50 & \cellcolor[rgb]{0.80,0.80,0.80} 39.35\\
\bottomrule

\end{tabular}
\vspace{-2mm}
\end{table}

\subsection{Error Analysis}

\begin{figure}[t]
    \centering
    \subfloat[\footnotesize \textnormal{M1 on VQ Task}]{%
        \includegraphics[width=0.32 \linewidth]{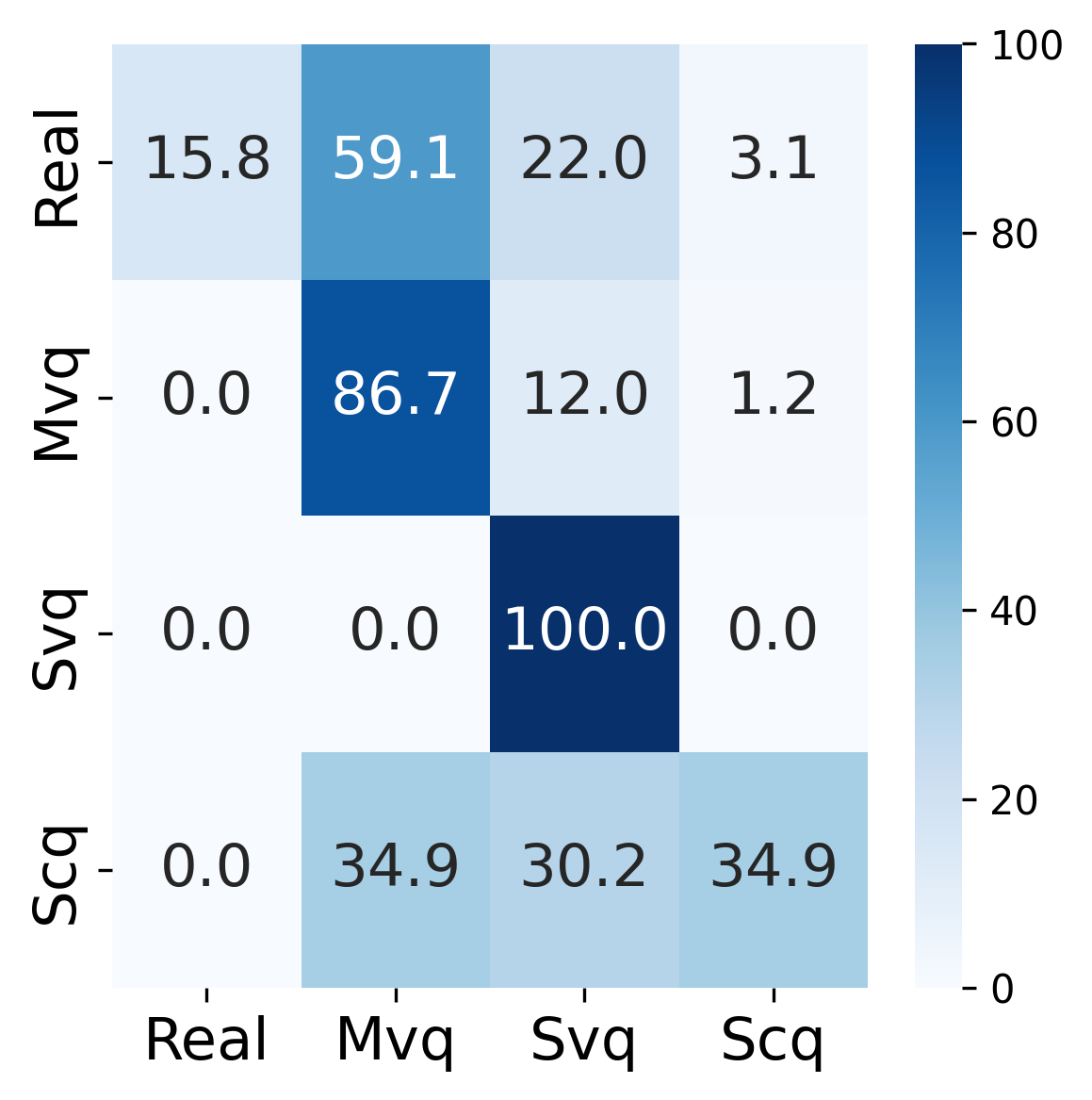}%
        \label{fig:m1_vq}
    }
    \hfill
    \subfloat[\footnotesize \textnormal{M1 on AUX Task}]{%
        \includegraphics[width=0.32 \linewidth]{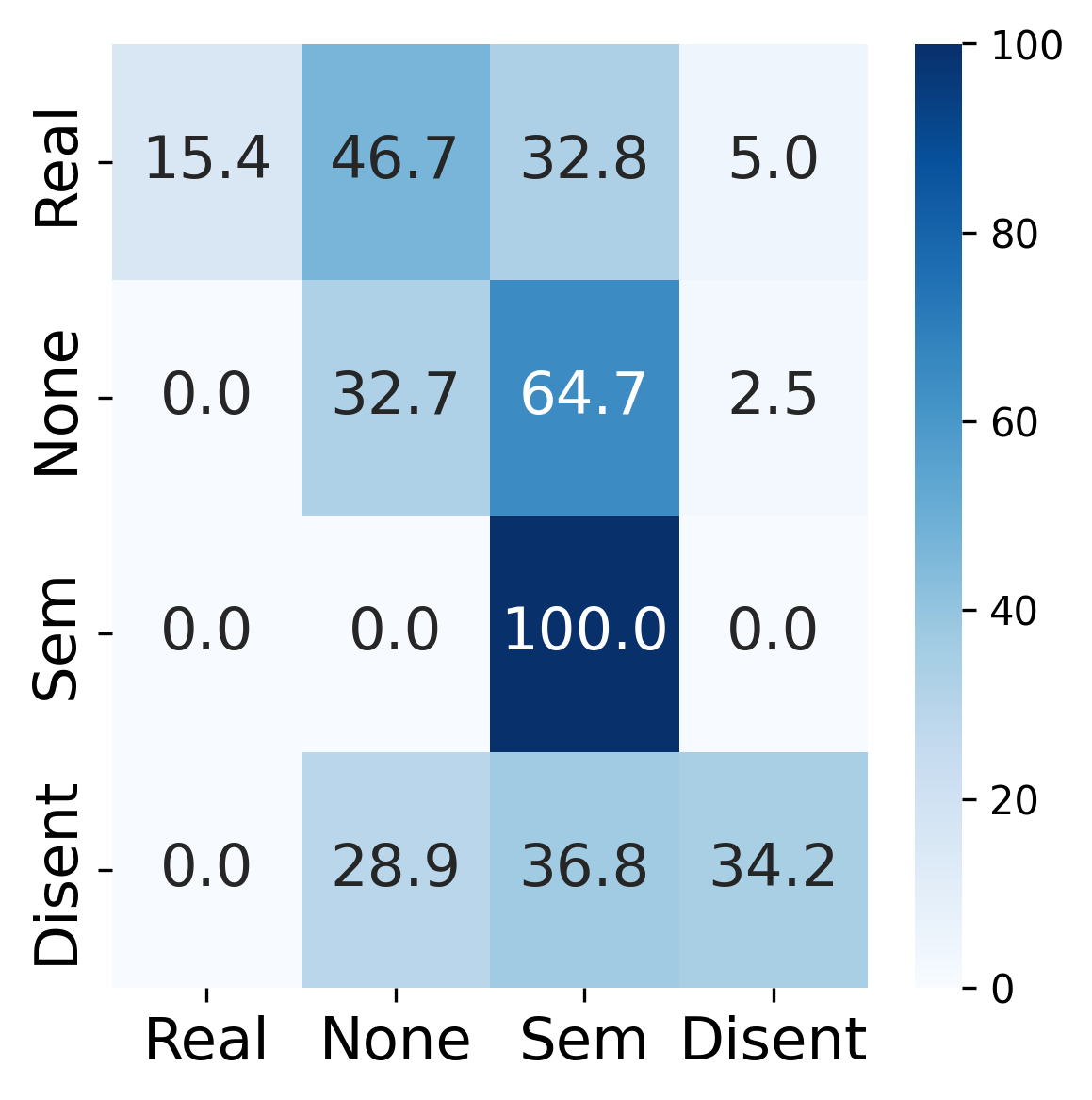}%
        \label{fig:m1_aux}
    }
    \hfill
    \subfloat[\footnotesize \textnormal{M1 on DEC Task}]{%
        \includegraphics[width=0.32 \linewidth]{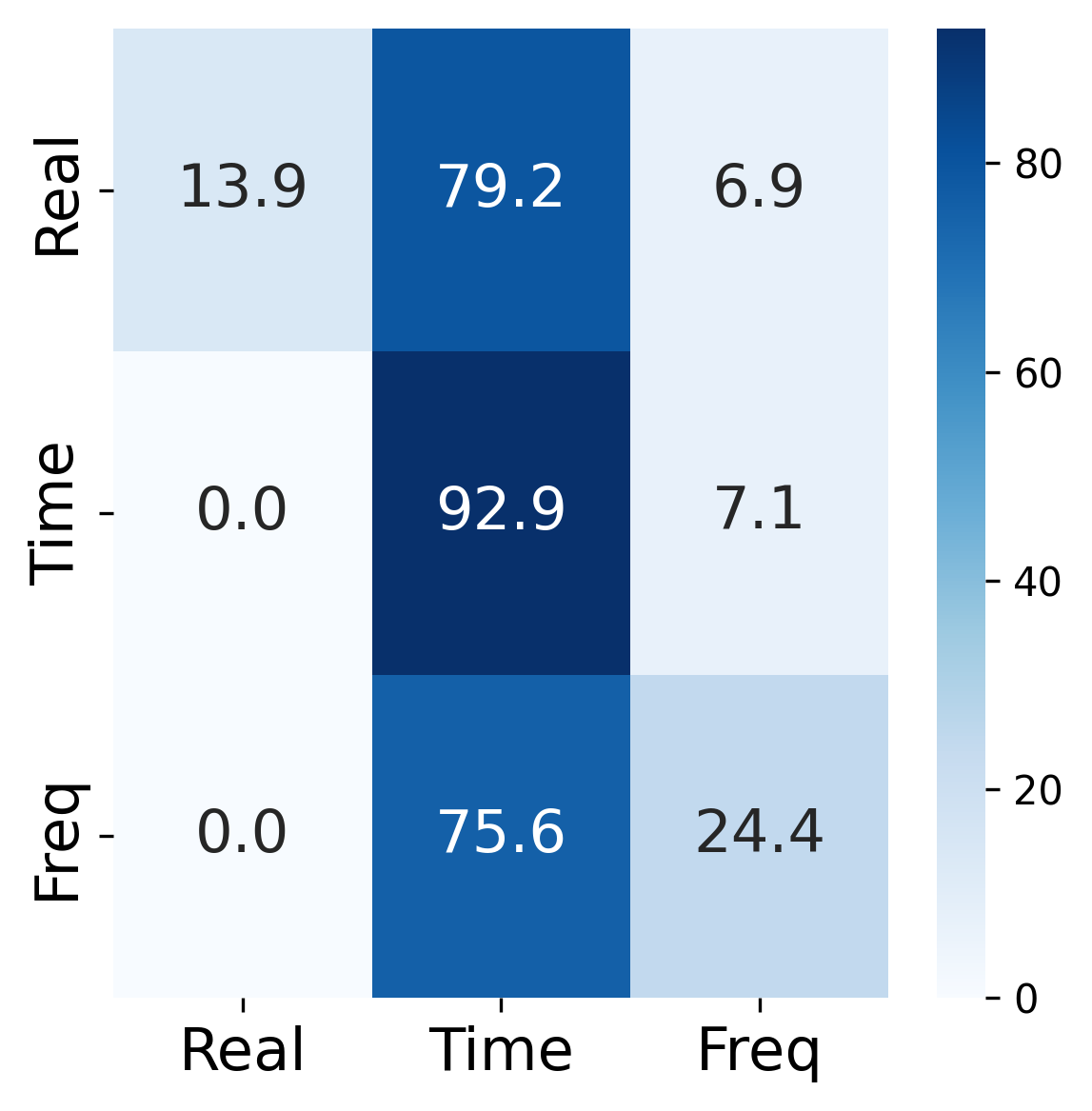}%
        \label{fig:m1_dec}
    }
    \vspace{-1em}
    \subfloat[\footnotesize \textnormal{M2 on VQ Task}]{%
        \includegraphics[width=0.32 \linewidth]{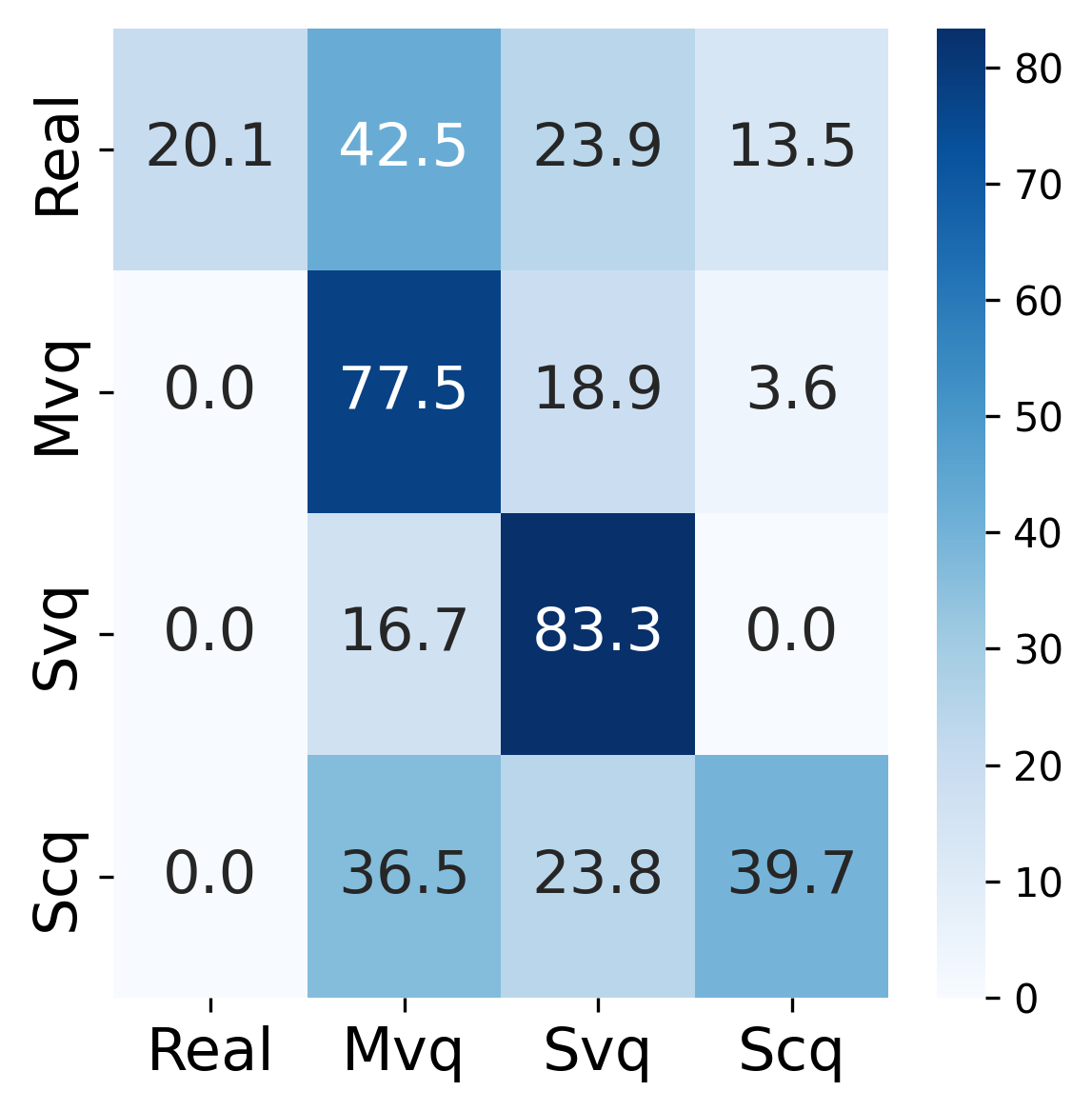}%
        \label{fig:m2_vq}
    }
    \hfill
    \subfloat[\footnotesize \textnormal{M2 on AUX Task}]{%
        \includegraphics[width=0.32 \linewidth]{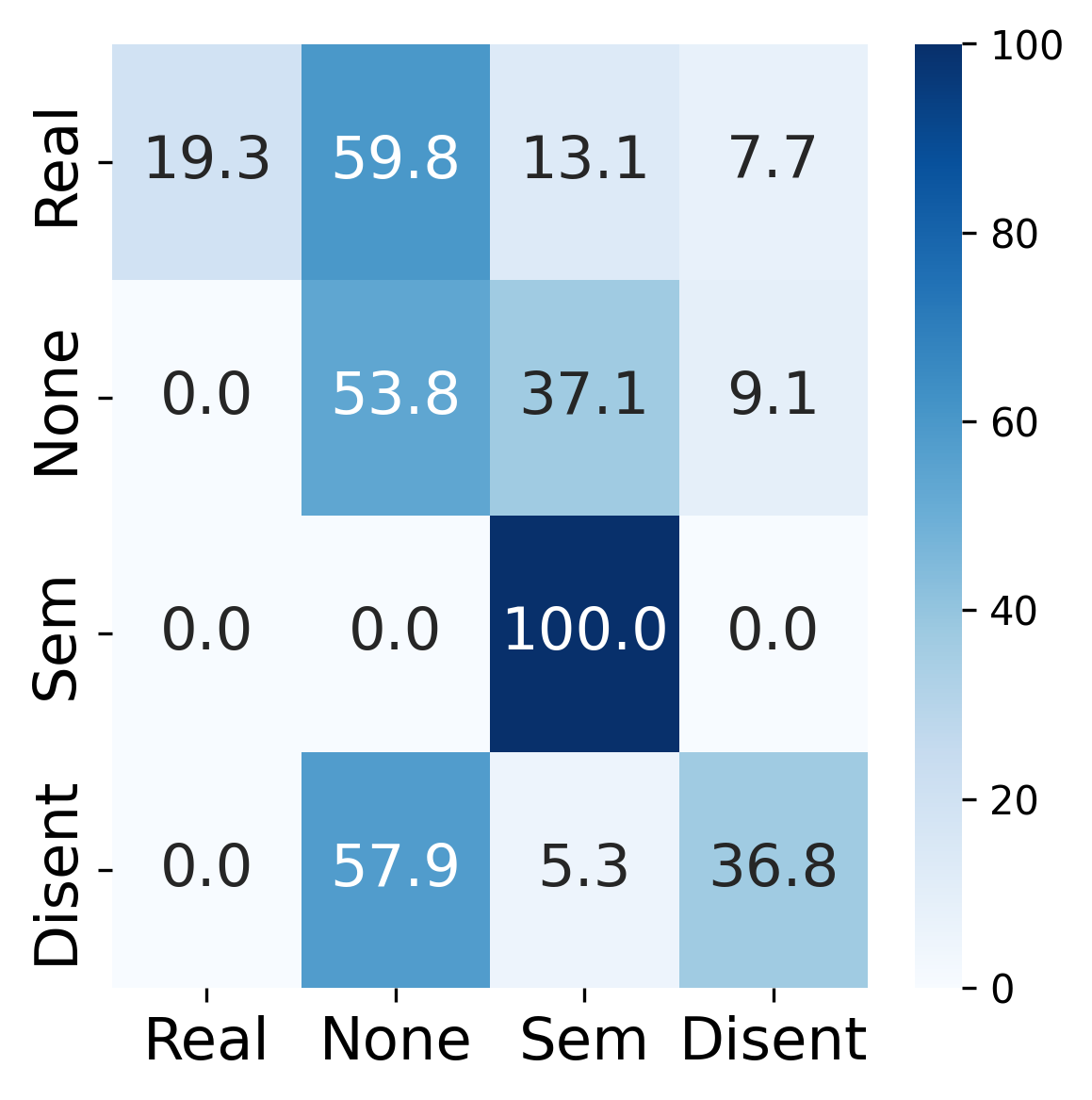}%
        \label{fig:m2_aux}
    }
    \hfill
    \subfloat[\footnotesize \textnormal{M2 on DEC Task}]{%
        \includegraphics[width=0.32 \linewidth]{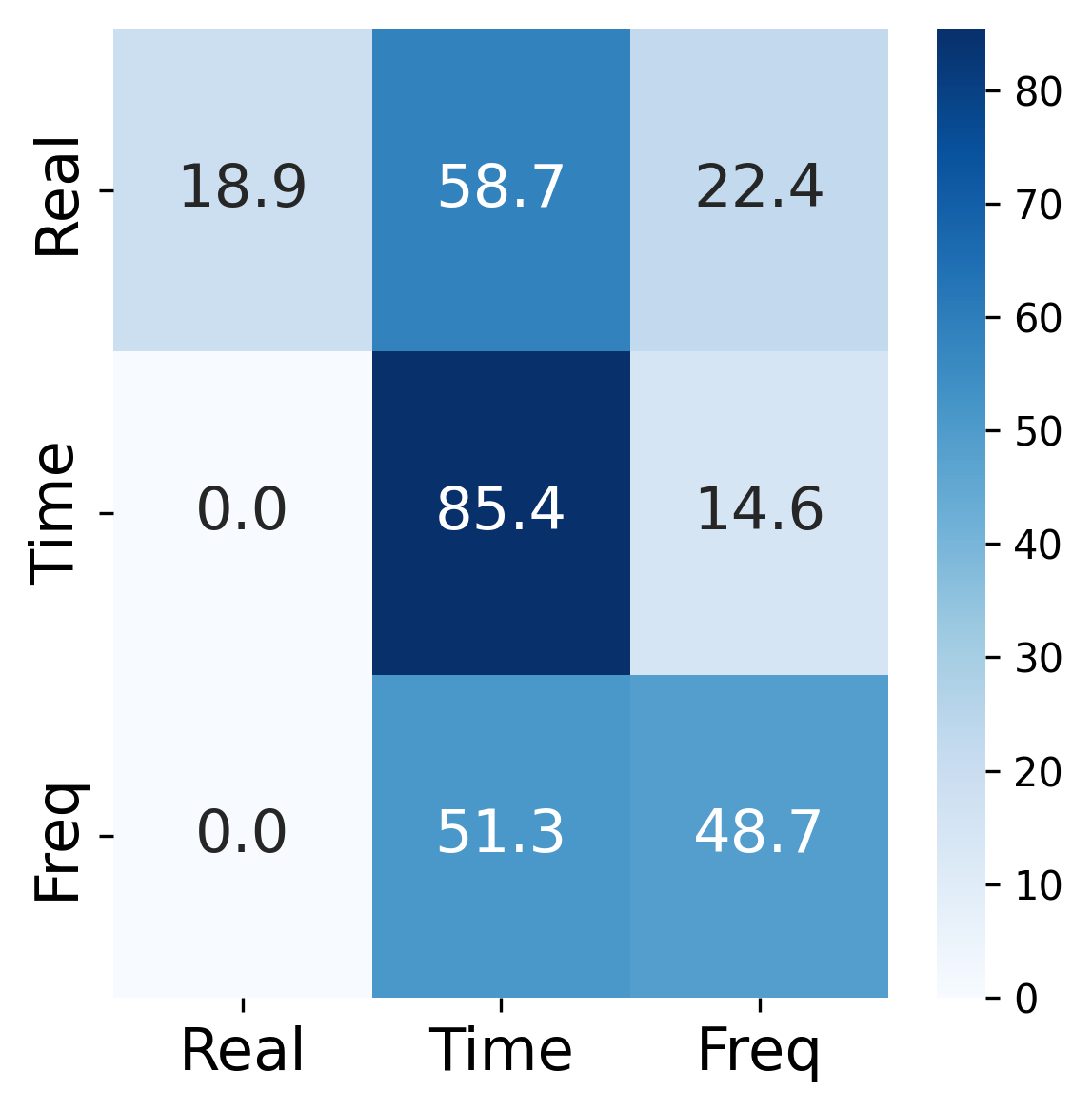}%
        \label{fig:m2_dec}
    }
        \vspace{-0.5em}
    \caption{Confusion matrices of M1 and M2 on CoSG (kn. codec) subset, row-normalized by true labels, with predictions on the horizontal axis and true labels on the vertical axis.}
    \vspace{-3mm}
    \label{fig:impact_multi_task}
\end{figure}

In the previous section, we observed that M2, trained on the AUX-balanced dataset, achieves the best generalization performance on source tracing. 
To further investigate, we compared the confusion matrices of model M1 and M2 (both trained on the AUX-balanced dataset) evaluation on different source tracing tasks as shown in Figure \ref{fig:impact_multi_task}. 
M1 frequently misclassifies bona fide speech, whereas M2 correctly identifies more bona fide samples, significantly improving overall performance.

One reason M1 underperforms M2 is its multitask approach, which must handle both bona fide/spoof classification and multi-class classification for source tracing, and can often be confusing. Meanwhile, M2 focuses solely on source tracing, leading to more consistent performance. Another factor is a domain mismatch in bona fide speech data, which leads to misclassifying bona fide speech as a certain category in the codec taxonomy and diminishes overall generalization. 
Another study \cite{xie2025neural} also noted that unseen real speech significantly reduced source tracing performance in codec identification.

\section{Conclusion}

This paper defines source tracing for deepfakes generated by codec-based speech generation models. We introduce three taxonomy-guided multi-class classification tasks: vector quantization, auxiliary objectives, and decoder type. 
Building on our defined source tracing framework, we compare different training strategies, including single-task, dual-task, and multi-task learning. We also explore taxonomy-guided data balance selection. Results on the \textit{CodecFake+} dataset show that source tracing models are susceptible to training data distributions.

As general-purpose audio-language models become central to speech understanding~\cite{huang2025dynamicsuperb}, many rely on them as default tools. 
Yet even the leading models like GPT-4o may decline deepfake detection requests due to built-in safety constraints \cite{lin2025preliminary}. 
In contrast, leveraging a neural audio codecs taxonomy, our source-tracing framework enables precise analysis of codec-based speech synthesis. We anticipate that this work will yield valuable insights and inspire further research.

\section{Acknowledgements}
This work was partially supported by the National Science and Technology Council, Taiwan (Grant no. NSTC 112-2634-F-002-005, Advanced Technologies for Designing Trustable AI Services). We also thank the National Center for High-performance Computing (NCHC) of National Applied Research Laboratories (NARLabs) in Taiwan for providing computational and storage resources.

\bibliographystyle{IEEEtran}
\bibliography{refs}

\end{document}